\documentclass[pra,aps,preprint,epsfig,showkeys,showpacs,superscriptaddress]{revtex4}
\usepackage{amsmath}
\usepackage{amssymb}
\usepackage[dvips]{graphicx}%\usepackage{graphicx}% Include figure files
\usepackage{dcolumn}% Align table columns on decimal point
\usepackage{bm}% bold math
\usepackage[colorlinks=false,dvipdfm]{hyperref} % citecolor=blue 文件内部引用
\usepackage{setspace}
\usepackage{amsfonts}
\usepackage{rotating}
\usepackage{makeidx}
\usepackage{amsthm}

\begin{document}

\title{Prepotential approach to systems with dynamical symmetries}
\author{Yan Li}
\affiliation{Theoretical Physics Division, Chern Institute of Mathematics, Nankai University, Tianjin 300071,
China}
\author{Fu-Lin Zhang}
\email{flzhang@tju.edu.cn}
\affiliation{Physics Department, School of Science, Tianjin
University, Tianjin 300072, China}
\author{Jing-Ling Chen}
\email{chenjl@nankai.edu.cn}
\affiliation{Theoretical Physics Division, Chern Institute of
Mathematics, Nankai University, Tianjin 300071, 
China}
\affiliation{Centre for Quantum Technologies, National University of Singapore, 3 Science Drive 2, Singapore 117543}
\author{L. C. Kwek}
\affiliation{Centre for Quantum Technologies, National University of Singapore, 3 Science Drive 2, Singapore 117543}
\affiliation{National Institute of Education and Institute of Advanced Studies, Nanyang Technological University, 1 Nanyang Walk, Singapore 637616}

\begin{abstract}
A prepotential approach to constructing the quantum systems with dynamical
symmetry is proposed. As applications, we derive generalizations
of the  hydrogen atom and harmonic oscillator, which can be regarded
as the systems with position-dependent mass. They have the symmetries
which are similar to the corresponding ones, and can be solved by
using the algebraic method. 
\end{abstract}

 \maketitle

\section {Introduction}\label{intro}

It is known that the solvable models provide essential
insights in the understandings of physical theory.
Much effort has been made to find solvable models \cite{Bertrand,carinena2004one,carinena2007quantum,mathews1974unique,lakshmanan2003nonlinear,nishijima1972green,higgs1979dynamical}, and,
in the meanwhile, various methods have also been proposed to solve systems,
such as, the raising and the lowing operators method \cite{wang2009raising},
factorization method \cite{wang2011factorization},
supersymmetric approach \cite{gonul2002supersymmetric}.
In this process, symmetries have been found to paly a fundamental role in the
solvable systems, and  knowledge of their presence in certain problems
often simplifies the solutions considerably. All
the well-known exactly solvable models as well as the quasi-exactly
solvable models have specific symmetries in quantum mechanics.
Among them, the hydrogen atom and the harmonic oscillator are the most
important for their authenticity and simplicity. The dynamical symmetry of the $N$-dimensional hydrogen atom is described
by the $SO(N+1)$ Lie group \cite{fock1935theorie,cisneros1969symmetry}, and the $N$-dimensional harmonic oscillator is shown to have
 the
$SU(N)$ dynamical symmetry \cite{Jauch,elliott1958collective}.

 For our purpose,
we would like to review  some facts on the symmetries of
 the hydrogen atom and the harmonic oscillator.
It is known that the classical orbits of these two systems are closed, and
the conserved quantities responsible for the closeness of classical orbits
have been proven to be the Rung-Lenz vector \cite{Lenz,Pauli} and the second order
tensors \cite{fradkin1965} respectively. The symmetries are called dynamical symmetries because
the nature of them is not geometrical but the symmetries in the phase space.
These symmetries lead to
an algebraic approach to determine the energy levels.

In this paper, we analyze
the symmetries of the hydrogen atom and the harmonic oscillator, and
try to  propose some new quantum systems with dynamical symmetries.
As the main contribution, we show that exactly solvable systems that follow
specific dynamical symmetries can be constructed in a rather simple way.
The  idea is that we  construct new generators of symmetric group from the
known ones and get the corresponding Hamiltonian,
which allow us to obtain exactly solvable systems.
 We call this constructing method as  prepotential approach, which is motivated by the article \cite{ho2008prepotential}. The generators we supposed also contain undetermined functions. They are called the prepotential, a concept
which plays a fundamental role in supersymmetric quantum mechanics \cite{cooper1995supersymmetry,junker1996supersymmetric}.

This paper is organized as follows. In Section 2, we give
a description of the prepotential approach. In Section 3, we present
an explicit construction of a Coulomb-like system.
We first construct the system in the 2-dimensional space. We then extend to
the 3-dimensional space and in general to the $N$-dimensional space.
In Section 4, we construct a 2-dimensional oscillator-like
system and generalize
it to the $N$-dimensional case.
In the final section, we conclude with summary and some perspective.

\section{Prepotential approach}\label{lilun}

In this section, we will describe  the idea that we use to
 construct some systems from  the symmetry  algebras directly.
In fact, we illustrate this idea by considering  the central potential in 2-dimensional
  space, and  then an extension  to the general $N$-dimension.

  The central potential
   in 2-dimensional space have the conserved  physical quantities:
   the angular momentum $L_{12}$ (In general, the angular momentum is denoted
   $L_{ij}=x_ip_j-x_jp_i$; and $L_{12}$ is the unique one  in 2-dimensional space.) and the Hamiltonian $ H$.
   Since $H$ and $L_{12}$ obey that
\begin{eqnarray}
\left[L_{12},H\right]=0,
\end{eqnarray}
it follows that they have the common eigenstates. Let $|E,m\rangle$ be the eigenstates to  eigenenergy $E$ and the angular momentum eigenvalue $m$. We have
\begin{eqnarray}
H|E,m\rangle=E|E,m\rangle, \ \ L_{12}|E,m\rangle=m|E,m\rangle.
\end{eqnarray}
 It is quite common that the Hamiltonian should be  of the form
\[H=\frac{1}{4M(r)}p^2+p^2\frac{1}{4M(r)}+h(L_{12})+V(r),\]
where $M(r)$ is position-dependent effective mass and $h(L_{12})$ is an arbitrary function on $L_{12}$.

Motivated by the algebraic structure  of the 2-dimensional hydrogen atom,
we make the assumption that, besides the  angular momentum $L_{12}$,
 there are some other physical quantities $T_{\pm}$
(In order to ensure the hermiticity, it is  required that  $T_+=(T_{-})^\dag$.)
that satisfy  the following  commutation relations
\begin{eqnarray}\label{duiyiguanxi}
\left[L_{12}, T_{\pm}\right]=\pm T_{\pm},\ \
\left[T_+,T_-\right]=F(H,L_{12})
\end{eqnarray}
as well as  the anticommutation relation
\begin{eqnarray}\label{fanduiyi}
\{T_+, T_-\}=G(H,L_{12}),
\end{eqnarray}
where $F$ and $G$ are arbitrary functions on $H$ and $L_{12}$.
This enables  us to construct systems with Lie algebra symmetries.

In view of (\ref{duiyiguanxi}) and (\ref{fanduiyi}), it is immediate  that
\begin{eqnarray}
T_+T_-=\frac{1}{2}(G(H,L_{12})+F(G,L_{12})),\\
T_-T_+=\frac{1}{2}(G(H,L_{12})-F(G,L_{12})).\nonumber
\end{eqnarray}
So,
\begin{eqnarray}
\langle E,m|T_+T_-|E,m\rangle=\frac{1}{2}(G(E,m)+F(E,m)),\\
\langle E,m|T_-T_+|E,m\rangle=\frac{1}{2}(G(E,m)-F(E,m)).\nonumber
\end{eqnarray}
Besides this, we know
\begin{eqnarray}\label{+_+}
T_+T_-T_+|E,m\rangle=T_+(T_-T_+)|E,m\rangle=(T_+T_-)T_+|E,m\rangle,\\
T_-T_+T_-|E,m\rangle=T_-(T_+T_-)|E,m\rangle=(T_-T_+)T_-|E,m\rangle,\nonumber
\end{eqnarray}
In the above equations, we can see that the states $T_+|E,m\rangle$ and $T_-|E,m\rangle$ are
the common eigenstates of $L_{12}$ and $H$.
Therefore, we obtain that
\begin{eqnarray}\label{R+R-}
T_+|E,m\rangle=\sqrt{\frac{1}{2}\left(G(E,m)-F(E,m)\right)}|E',m+1\rangle,\\
T_-|E,m\rangle=\sqrt{\frac{1}{2}\left(G(E,m)+F(E,m)\right)}|E'',m-1\rangle,\nonumber
\end{eqnarray}

The $L_{12}$ span a finite-dimensional
subspace.
There exist a highest and a lowest weights for its representation, denoted by
\begin{eqnarray}
T_+|E,\overline{m}\rangle=0, \ \ \ T_-|E,\underline{m}\rangle=0.
\end{eqnarray}
Therefore, from (\ref{R+R-}), we know
\begin{eqnarray}\label{9}
G(E,\overline{m})-F(E,\overline{m})=0\ \ \
\end{eqnarray}
and
\begin{eqnarray}\label{8}
G(E,\underline{m})+F(E,\underline{m})=0.
\end{eqnarray}
The degeneracy of states in this representation space should be a natural number, which leads to $\overline{m}-\underline{m}=n=0,1,2\ldots$.

%Applying  equations (\ref{R+R-}) and (\ref{+_+}), one can
%obtain the following recurrence relations
%\begin{eqnarray}\label{10}
%G(E',m+1)+F(E',m+1)=G(E,m)-F(E,m),\\
%G(E,m)+F(E,m)=G(E'',m-1)-F(E'',m-1).\nonumber
%\end{eqnarray}
%Moreover, from (\ref{R+R-}) we know that
%\begin{eqnarray}\label{9}
%G(E,\overline{m})-F(E,\overline{m})=0
%\end{eqnarray}
%for the upper bound of the states $|E,\overline{m}\rangle$, and
%\begin{eqnarray}\label{8}
%G(E,\underline{m})+F(E,\underline{m})=0
%\end{eqnarray}
%for the lower bound $|E,\underline{m}\rangle$.
%Thus we obtain the energy for the maximum  and minimum eigenvalues
%of angular momentums $\overline{m}$ and $\underline{m}$.
%****Other energies of the angular momentum  $L_{1,2}$
%are gained by the recurrence relations (\ref{10}).*****

The Hamiltonian can be  obtained  by comparing  Eqs. (\ref{duiyiguanxi}) and (\ref{fanduiyi}) and adjusting the form of the $T_\pm$. The key ingredient of  our approach  is to choose suitable $T_{\pm}$.
When $T_{\pm}$ are given,  we can  get the eigenenergy spectrum from Eqs. (\ref{9}) and (\ref{8}).
It should be emphasized that  $ T_{\pm}$ are not required to be conserved quantities. While, we get the common eigenstates of $L_{12}$ and $H$ when
they act on the state $T_+|E,m\rangle$ and $T_-|E,m\rangle$.
The above idea will be  demonstrated with concrete  examples in Sections 3 and 4, where we will construct a Coulomb-like system and a oscillator-like system respectively.

\section{Coulomb-like system}

The purpose of this section  is to construct a system analogous to hydrogen atom by using the method described in the preceding section. We now give a brief description of the strategy that is used  to construct the new system $R_i$ $(i=1,2)$  from the original system.

 In the hydrogen atom system, the two components of Runge-Lenz vector $$R_1=\frac{1}{2}(p_2L_{12}+L_{12}p_2)-\frac{x_1}{r}, \ \ \ \ \ R_2=-\frac{1}{2}(p_1L_{12}+L_{12}p_1)-\frac{x_2}{r}$$
  satisfy the commutation relations (in unit $\hbar=1$)
\begin{eqnarray}
\left[L_{12}, R_{\pm}\right]=\pm R_{\pm},\ \ \
\left[R_+,R_-\right]=-4HL_{12}
\end{eqnarray}
and the anticommutation relation
\begin{eqnarray}
\{R_+, R_-\}=(4L^2+1)H+1,
\end{eqnarray}
where $R_{\pm}=R_1\pm iR_2$, and the scalar $L$ is defined as $L^2=L_{ij}L_{ij}$ .

We wish to construct new physical quantities that satisfy the following relations
\begin{eqnarray}\label{afj}
\left[L_{12}, R_{\pm}\right]=\pm R_{\pm},\ \
\left[R_+,R_-\right]=F(H,L_{12})
\end{eqnarray}
and the anticommutation relation
\begin{eqnarray}\label{3}
\{R_+, R_-\}=G(H,L_{12}).
\end{eqnarray}

To imitate the hydrogen atom, we assume that
\begin{eqnarray}\label{rr}
R_1=\frac{1}{2}(f(r)p_2L_{12}+L_{12}p_2f(r))+g(r)x_1,\ \ \ \ \ \
R_2=-\frac{1}{2}(f(r)p_1L_{12}+L_{12}p_1f(r))+g(r)x_2,
\end{eqnarray}
which satisfy
$\left[L_{12}, R_1\right]=iR_2, \left[L_{12},R_2\right]=-iR_1$ (i.e. $\left[L_{12}, R_{\pm}\right]=\pm R_{\pm}$). The undetermined functions $f(r)$ and $g(r)$ are the prepotential.
The anticommutation relation can be easily obtained
\begin{eqnarray}
\{R_+, R_-\}&=&2R_1^2+2R_2^2\nonumber\\
&=&\frac{1}{2}\left(f(r)pf(r)p+pf^2(r)p+f(r)p^2f(r)+pf(r)pf(r)+8f(r)g(r)\right)L_{12}^2
+2g^2(r)r^2.
\end{eqnarray}
In order to get the relation (\ref{3}), we see that $g^2(r)r^2$ must be a constant. For simplicity and consistency with the hydrogen atom, we select
\begin{eqnarray}
g(r)=-\frac{1}{r}.
\end{eqnarray}

For brevity, we introduce a new operator $\pi_i$. Such an idea has also appeared in  the paper \cite{higgs1979dynamical}. Since $\pi_i$ is the hermite operator, it can be written as
\begin{eqnarray}
\pi_1=\frac{1}{2}(f(r)p_1+p_1f(r)), \ \ \pi_2=\frac{1}{2}(f(r)p_2+p_2f(r)).
\end{eqnarray}
So $R_1$ and  $R_2$ can be written as
\begin{eqnarray}\label{RR}
R_1=\frac{1}{2}(\pi_2L_{12}+L_{12}\pi_2)-\frac{x_1}{r},\ \ \ R_2=-\frac{1}{2}(\pi_1L_{12}+L_{12}\pi_1)-\frac{x_2}{r}.
\end{eqnarray}
Equations (\ref{rr}) and (\ref{RR}) do not coincide. This is because
(\ref{rr}) is the semiclassical attempt. In order to keep the symmetry and the hermiticity, we use the description (\ref{RR}).

Because $\left[\pi_1,\pi_2\right]=-i\frac{f(r)f'(r)}{r}L_{12}$, we have
\begin{eqnarray}\label{aa}
\left[R_1, R_2\right]=-2iL_{12}\left(\frac{\pi^2}{2}-\frac{f(r)}{r}+\frac{f(r)f'(r)}{2r}L^2\right).
\end{eqnarray}
The expression in the bracket on the right-hand side of (\ref{aa})
 is a function of $H$ and $L_{12}$. So, we can take $f(r)f'(r)=\lambda r$,
where $\lambda$ is an arbitrary parameter. Then it follows  that  $f(r)=\sqrt{C+\lambda r^2}$, where $C$ is an arbitrary constant.
 If we wish that  the system reduces  to the original hydrogen atom as $\lambda\rightarrow0$,
then we have
\begin{eqnarray}
f(r)=\sqrt{1+\lambda r^2}.
\end{eqnarray}

Now, we see that (\ref{aa}) has the following form
\begin{eqnarray}
\left[R_1,R_2\right]=-2iL_{12}\left(\frac{\pi^2}{2}+\frac{\lambda L^2}{2}-\frac{\sqrt{1+\lambda r^2}}{r}\right)
\end{eqnarray}
and the length of Rung-Lenz vector can be written as
\begin{eqnarray}
R_1^2+R_2^2=2L^2\left(\frac{\pi^2}{2}-\frac{\sqrt{1+\lambda r^2}}{r}\right)+\frac{1}{2}\left(\frac{\pi^2}{2}-\frac{\sqrt{1+\lambda r^2}}{r}\right)+\lambda L^2+1.\nonumber
\end{eqnarray}

In free particle motion, we require conservation of linear momentum $p_i$ is replaced by
conservation of the vector $\pi_i$. This means the vector $\pi_i$ is commute with the Hamiltonian
when the potential vanishes.
Therefore, the Hamiltonian of our constructed system is
\begin{eqnarray}\label{Hamiltonian}
H=\frac{\pi^2}{2}-\frac{\lambda L^2}{2}-\frac{\sqrt{1+\lambda r^2}}{r},
\end{eqnarray}
where $\frac{\pi^2}{2}-\frac{\lambda L^2}{2}$ is the term for kinetic energy, and the remaining  is the  potential. Note that this system can be considered as a hydrogen atom with
a position-dependent effective mass $M(r)=(1+\lambda r^2)^{-1}$.

In view of (\ref{afj}) and (\ref{3}), we get
\begin{eqnarray}\label{r-1}
F(H,L_{12})=-4L_{12}(H+\lambda L^2)
\end{eqnarray}
and
\begin{eqnarray}\label{r-2}
G(H,L_{12})=(2L^2+\frac{1}{2})(2H+\lambda L^2)+2\lambda L^2+2.
\end{eqnarray}
Substituting (\ref{r-1}) and (\ref{r-2}) into (\ref{9}) and (\ref{8})
 and by a straightforward calculation, the eigenenergy spectrum of the system (\ref{Hamiltonian}) is
\begin{eqnarray}
E_{n}=-\frac{2}{(2n+1)^2}-\frac{1}{2}\lambda n(n+1)-2\lambda n(n+1)\frac{2n+2}{(2n+1)^2}.
\end{eqnarray}

Now we   extend the above  to the 3-dimensional situation, which is more difficult  than the 2-dimensional case.
The new constructed physical quantity is a vector, which can be written as
\begin{eqnarray}
\vec{R}=\frac{1}{2}(\vec{\pi}\times\vec{L}-\vec{L}\times\vec{\pi})-\frac{\vec{r}}{r},
\end{eqnarray}
 where $\vec{L}$ is the vector ($L_{1}, L_{2}, L_{3}$)=($L_{23}, L_{31}, L_{12}$) and $\vec{\pi}$ is the vector ($\pi_1, \pi_2, \pi_3$) with $\pi_i=\frac{1}{2}\left(\sqrt{1+\lambda r^2}p_i+p_i\sqrt{1+\lambda r^2}\right)$ $(i=1,2,3)$.

 Noting that
$$
\left[L_\alpha, L_\beta\right]=i\epsilon_{\alpha\beta\gamma}L_{\gamma},\ \ \left[L_\alpha, \pi_\beta\right]=i\epsilon_{\alpha\beta\gamma}L\pi_{\gamma}, \ \left[L_\alpha, R_\beta\right]=i\epsilon_{\alpha\beta\gamma}R_{\gamma} \ \ \ \ \ \alpha,\beta,\gamma=1,2,3$$
through a complicated computation, we obtain
\begin{eqnarray}
\vec{R}\times\vec{R}%&=&\left[\sqrt{1+\lambda r^2}(p^2\vec{r}-(\vec{p}\cdot\vec{r})\vec{p})-\frac{1}{2}i\lambda\frac{1}{\sqrt{1+\lambda r^2}}((\vec{r}\cdot\vec{p})\vec{r}-r^2\vec{p})-\frac{\vec{r}}{r}\right]\nonumber\\
%&&\times\left[(\vec{r}p^2-3i\vec{p}-\vec{p}(\vec{p}\cdot\vec{r}))\sqrt{1+\lambda r^2}-\frac{1}{2}i\lambda(\vec{r}(\vec{r}\cdot\vec{p})-\vec{p}r^2-3i\vec{r})\frac{1}{\sqrt{1+\lambda r^2}}-\frac{\vec{r}}{r}\right]\nonumber\\
&=&-2i(H+\lambda L^2)\vec{L}.
\end{eqnarray}
On the other hand, it is not hard to get the following
\begin{eqnarray}
\vec{R}\cdot\vec{R}=(2H+\lambda L^2)L^2+2H+2\lambda L^2+1.
\end{eqnarray}
The energy spectrum is similar to the 2-dimensional case, and it is written as
\begin{eqnarray}
E_{n}=-\frac{1}{2(n+1)^2}-\frac{1}{2}\lambda n(n+2)-\lambda n(n+2)\frac{n+1}{(n+1)^2}.
\end{eqnarray}

In fact, we can also extend to
the N-dimensional Coulomb-like system. It turns out that
\begin{eqnarray}
\left[L_{ij}, R_k\right]&=&i(\delta_{ik}R_{j}-\delta_{jk}R_i),\ \ \ \ \ \ \ \ \ \ \ \left[R_i,R_j\right]=-2i(H+\lambda L^2)L_{ij},  \\
R_iR_i&=&(2H+\lambda L^2)L^2+\frac{(N-1)^2}{2}(H+\frac{\lambda L^2}{2})+\lambda L^2+1,\nonumber\\
E_{n}&=&-\frac{1}{2(n+\frac{N-1}{2})^2}-\frac{1}{2}\lambda n(n+N-1)-\lambda n(n+N-1)\frac{n+1}{(n+\frac{N-1}{2})^2},\nonumber
\end{eqnarray}
where $R_i=\frac{1}{2}(L_{ij}\pi_j-\pi_jL_{ij})-\frac{x_i}{r}$ is the conserved quantity.

Now we have completed the construction for the Coulomb-like system whose Hamiltonian is (\ref{Hamiltonian}). Here, we would like to remark that
it seems to be rather difficult to
construct high-dimensional systems by directly using the prepotential approach because of the numerous amount of conserved quantities. However, as illustrated in this section,  it is feasible to
construct high-dimensional systems by first establishing the 2-dimensional case.

\section{Oscillator-like system}\label{xiezhenzi}

In this section, we  turn to the construction of  oscillator-like system. To this end, we need to construct the physical quantities $Q_{xy}$ and $Q_1$ which satisfy the following  commutation relations
\begin{eqnarray}\label{5}
\left[Q_{xy},L_{12}\right]=2iQ_1,\ \ \ \ \ \ \ \left[Q_1,L_{12}\right]=-2iQ_{xy},\ \ \ \ \ \ \
\left[Q_{xy},Q_1\right]=F(H,L_{ij}),
\end{eqnarray}
and the anticommutation relation
\begin{eqnarray}\label{4}
Q_{xy}^2+Q_1^2=G(H,L_{ij}).
\end{eqnarray}
Analogous  to the harmonic oscillator, we  assume that
\begin{eqnarray}\label{33}
Q_{xy}&=&\frac{1}{2}\left(f(r)p_1p_2+p_1p_2f(r)\right)+g(r)x_1x_2,\nonumber\\
Q_{1}&=&\frac{1}{2}g(r)(x_1^2-x_2^2)+\frac{1}{4}(f(r)p_1^2+p_1^2f(r)-f(r)p_2^2-p_2^2f(r)).
\end{eqnarray}

Now we present the construction similar to
the Coulomb-like system. By a
direct computation, there holds that
\begin{eqnarray}\label{7}
Q_{xy}^2+Q_{1}^2=\frac{1}{16}\left(f^2(r)p^2+p^2f^2(r)\right)+\frac{1}{4}g^2(r)r^2,\\
\label{6}\left[Q_{xy},Q_{1}\right]=f(r)g(r)L_{12}+\frac{1}{4}f(r)G(x_1,p_1,x_2,p_2),
\end{eqnarray}
where $G(x_1,p_1,x_2,p_2)$ is a function of $x_1,p_1,x_2,p_2$,
and is related to the function $f(r)$.
We see that the right-hand
 side of equation (\ref{7})  must be a function of $H$ since
 $L_{12}$ does not appear  in it.
  In fact, because of the factor $p^2$, it is a reasonable assumption that
  it is just equal to
  $H$. Besides, we can see from  (\ref{6}) that $f(r)g(r)$ is a constant and $f(r)G(x_1,p_1,x_2,p_2)$ is a function of $H$ and $L_{12}$.

Inspired by Coulomb-like system, we wish to
obtain $Q_{x,y}$ and $Q_1$ of forms as given in (\ref{qq}).
For this purpose,  we choose $\pi_i=\frac{1}{2}\left(\sqrt{1+\lambda r^2}p_i+p_i\sqrt{1+\lambda r^2}\right)$
and thus
\begin{eqnarray}
f(r)=1+\lambda r^2\ \ \mbox{and}\ \ g(r)=\frac{1}{1+\lambda r^2}.
\end{eqnarray}

By the above analysis, we finish the constructions for the following quantities
\begin{eqnarray}\label{qq}
Q_{xy}&=&\frac{1}{2}\left(\pi_1\pi_2+\pi_2\pi_1\right)+\frac{x_1x_2}{1+\lambda r^2},\nonumber\\
Q_{1}&=&\frac{1}{2}\frac{x_1^2-x_2^2}{1+\lambda r^2}+\frac{1}{2}(\pi_1^2-\pi_2^2),
\end{eqnarray}
where the quantities $\vec{\pi}$ and $L_{ij}$ are the same as those for
hydrogen atom, namely,
\begin{eqnarray}
\pi_i&=&\frac{1}{2}(\sqrt{1+\lambda r^2}p_i+p_i\sqrt{1+\lambda r^2}),\\
L_{ij}&=&x_ip_j-x_jp_i.
\end{eqnarray}
These expressions are different from (\ref{33}), but they have no effect on the conclusion.

By a simple  calculation, it can be verified that
\begin{eqnarray}
\left[Q_{xy},L_{12}\right]&=&2iQ_1,\ \ \ \ \ \ \ \left[Q_1,L_{12}\right]=-2iQ_{xy},\nonumber\\
\left[Q_{xy},Q_1\right]&=&-2iL_{12}+\frac{1}{2}i\lambda^2L_{12}+i\lambda(2H+\lambda L^2)L_{12},
\end{eqnarray}
and
\begin{eqnarray}
Q_{xy}^2+Q_1^2&=&(H+\frac{\lambda L^2}{2})^2+\lambda H+\frac{5}{4}\lambda^2L^2-L^2-1.
\end{eqnarray}
Similar to the system of Coulomb-like potential, the Hamiltonian of oscillator-like system is
\begin{eqnarray}\label{Hamiltonian2}
H=\frac{\pi^2}{2}-\frac{\lambda L^2}{2}+\frac{\frac{1}{2}r^2}{1+\lambda r^2}.
\end{eqnarray}

We next consider an  extension to the N-dimensional case.  Because the conserved quantities are not suitable to generalize to the N-dimensional space, we
have to write them in another style. (In 2-dimensional case, we adopt the most common expression to  facilitate the reader.) Besides the conservative angular
momentum $L_{ij}$, we may define an extended
quadrupole tensor
\begin{eqnarray}
S_{ij}=\frac{1}{2}(\pi_i\pi_j+\pi_j\pi_i)+\frac{x_ix_j}{1+\lambda r^2}.
\end{eqnarray}
The components have the commutators
\begin{eqnarray}
\left[L_{ij},S_{kl}\right]=2i(-\delta_{jk}S_{il}-\delta_{jl}S_{ik}+\delta_{ik}S_{jl}+\delta_{il}S_{jk})
\end{eqnarray}
and
\begin{eqnarray}
\left[S_{ij},S_{kl}\right]=&&i\left[(1-\frac{1}{4}\lambda^2)(L_{ik}\delta_{jl}+L_{jk}\delta_{jl}+L_{jk}\delta_{il}+L_{jl}\delta_{ik})\right.\\
&&\left.-\frac{1}{2}\lambda(L_{ik}S_{jl}+L_{jk}S_{jl}+L_{jk}S_{il}+L_{jl}S_{ik})-\frac{1}{2}\lambda(S_{jl}L_{ik}+S_{jl}L_{jk}+S_{il}L_{jk}+S_{ik}L_{jl})\right].\nonumber
\end{eqnarray}
The scalars formed from them are
\begin{eqnarray}
I_1=S_{ii}=2H+\lambda L^2
\end{eqnarray}
and
\begin{eqnarray}
I_2=S_{ij}S_{ji}-S_{ii}S_{jj}=-2L^2-N(N-1)+\lambda \left[2(N-1)H-(N+\frac{1}{2})\lambda L^2\right].\nonumber
\end{eqnarray}
We can obtain the energy spectrum as given below
\begin{eqnarray}
E_{n}=(n+\frac{N}{2})\sqrt{1+\frac{1}{4}\lambda^2}-\frac{1}{2}\lambda(n^2+Nn+\frac{N}{2}).
\end{eqnarray}
In fact, this system has been %described in the paper \cite{carinena2004non}, and it is
studied by using both the
Lagrangian and the Hamiltonian formalisms in the paper \cite{carinena2004non}.
While, compared with their method, our
approach can give rise to the solution in a much simpler way.
Besides, in our approach the integral is also not required.

\section{Summary}

The prepotential approach to constructing models with dynamical symmetries is presented. We explicitly construct a Coulomb-like system and a oscillator-like system.  We hope that our approach could give rise to new systems that cannot be dealt with by the known methods.

In the paper, we present this method in central field, and we can generalize it to the non-central  field.
We rewrite the two generators of all the three in central field.  But in the non-central  field, as the angular momentum is no longer a special conservation, all the generators should be rewritten.  %In the following days, we will make an attempt.
Moreover, we assume that the prepotential approach can be popularized to the Dirac equation. Because the Dirac Hamiltonian is said to have the spin or pseudospin
symmetry corresponding to the same or opposite sign \cite{ginocchio2005relativistic}. When the potentials are spherical, the total angular momentum can be divided into conserved orbital and spin
parts, which form the $SU(2)$ algebra separately \cite{bell1975dirac,zhang2009higgs}.
We believe this approach can construct some other general classes of physical systems, especially the solvable models in quantum mechanics.

\acknowledgments
 This work is supported by the National Natural Science Foundation of China (Grant Nos. 11105097, 10975075 and 11175089), the
National Basic Research Program of China (Grant No. 2012CB921900), and the National Research Foundation and Ministry of
Education, Singapore.

\end{document}